\documentclass[12pt] {iopart}
\usepackage{graphicx}
\usepackage{amsfonts}

\begin{document}

\title[]{Copernicanism and the Typicality in Time}
\author{Milan M. \'Cirkovi\'c$^{1,2}$,  Amedeo Balbi$^{3}$}
\address{$^{1}$ Astronomical Observatory of Belgrade,
Volgina 7 11000 Belgrade, Serbia} 
\address{$^{2}$ Future of Humanity Institute, Faculty of Philosophy, University of Oxford,
Suite 8, Littlegate House, 16/17 St Ebbe's Street, Oxford, OX1 1PT, UK} 
\address{$^{3}$ Dipartimento di Fisica, Universit\`a di Roma ``Tor Vergata'', Via della Ricerca Scientifica, 00133 Roma, Italy}
\begin{abstract}
How special (or not) is the epoch we are living in? What is the appropriate reference class for embedding the observations made at the present time? How probable -- or else -- is anything we observe in the fulness of time? Contemporary cosmology and astrobiology bring those seemingly old-fashioned philosophical issues back into focus. There are several examples of contemporary research which use the assumption of typicality in time (or temporal Copernicanism) explicitly or implicitly, while not truly elaborating upon the meaning of this assumption. The present paper brings attention to the underlying and often uncritically accepted assumptions in these cases. It also aims to defend a more radical position that typicality in time is not -- and cannot ever be -- well-defined, in contrast to the typicality in space, and the typicality in various specific parameter spaces. This, of course, does not mean that we are atypical in time; instead, the notion of typicality in time is necessarily somewhat vague and restricted. In principle, it could be strengthened by further defining the relevant context, e.g., by referring to typicality within the Solar lifetime, or some similar restricting clause. 
\end{abstract}

\vspace{2pc}

\maketitle

\begin{quote}
I think we agree, the past is over. \\ --- George W. Bush (as reported by {\it Dallas Morning News}, May 10, 2000.)
\end{quote}

\section{Introduction}

The classical cosmological principle (henceforth CP) of Eddington and Milne enables us to make the notion of typicality in space intelligible -- at least on large, cosmological scale within the context of the Friedmann-Lema\^itre-Robertson-Walker family of models (henceforth FLRW). We may state that we are typical in space in virtue of accepting homogeneity and isotropy of the universe on large scales: each place is as good as any other. Of course, it is important to be careful about the meaning of ``place'' here, since the crucial step is to smooth out local features which would obviously make us atypical (we live on a planet, in a higher than average density region, with certain chemical abundances, etc.). Once we keep tabs on this, it makes sense to define an averaged, smoothed-out reference to which we can compare any spatial location. For quantities which depend on the choice of a proper system of reference in space, we may always switch to the system of reference associated with the cosmic microwave background (henceforth CMB). This has been excellently confirmed by the measurement of the proper motion of the Local Group of galaxies by observing the dipole anisotropy in CMB (e.g., Smoot, Gorenstein, and Muller 1977; Gorenstein and Smoot 1981). We can talk freely about the spatial typicality of any quantity or feature of the universe, since we have a well-defined average, offered in each hypersurface of constant cosmic time by the cosmological principle. For instance, we can meaningfully ask questions such as: \textit{Is the gaseous fraction of the baryonic mass of the Milky Way typical for all spiral galaxies? Is it typical that planets like Earth form near the co-rotation radius in spiral galaxies? Is the amount of chemical element X we perceive on Earth or in our Galactic vicinity somehow special or typical?} Etc. 


It is important to understand as precisely as possible what we are doing when posing and addressing such questions. We have a sample of natural phenomena, established by various empirical methods, which span a range in various parameter spaces, including 4-D spacetime. For example, galaxies form catalogues extended in space (and look-back time), but can also be represented as points in spaces of redshift or mass or morphological parameters, etc. Some of these measures are continuous, other discrete, but all are supposedly well-defined across the entire sample. There is an underlying (``natural'') distribution of these parameters; for example, CP indicates that the spatial distribution of galaxies or groups and clusters of galaxies is uniform on large scales, while in contrast morphological parameters are strongly concentrated toward those describing two main galaxy types. Underlying distributions are usually thought of as consequences of various cosmological and cosmogonical processes, but as we shall see below, this might be misleading when distributions in time are concerned.  
  
For spatial averages, CP serve us very well (this is not to say CP is entirely unproblematic: see, for instance, Beisbart (2009) for a cogent philosophical criticism).  Sufficiently large spatial volumes of the universe, with local fluctuations smoothed out, will be representative for the whole. If we claim, for instance, that the Sun is a typical star within a set of the Main Sequence stars in our Galaxy, there are well-defined criteria on the basis of which one could evaluate the claim (Gustafsson 1998; Robles et al. 2008); similarly for our Galaxy within a set of galaxies in a representative spatial volume of the universe (Hammer et al.\ 2007). Clearly, these issues are of key importance for astrobiology and our investigations of habitability of the universe (e.g., Chyba and Hand 2005; Gonzalez 2005). Copernicanism still serves as a powerful principle in astrobiology and many conclusions, such as those about the Galactic Habitable Zone, are impossible to reach without some Copernican assumption about the typicality. 

Now, we eminently do not have such a grounding principle for typicality in time, in spite of the relativistic embedding of time into cosmological spacetime and the key role of time in cosmology (e.g., Balbi 2013). With the demise of the classical steady state theory, in particular the version of Bondi and Gold (1948) based on the perfect cosmological principle (henceforth PCP), it has become clear that such a grounding principle is not only unavailable, but would contradict the very central notion of the evolving universe (Balashov 1994). PCP postulated homogeneity in both space and time, thus being a special case of CP with the highest level of symmetry. Obviously, PCP extremely constrained cosmological models, which has been regarded as a virtue of the steady-state theory in its heyday: there was just a single steady-state theory, compared with many possible Big Bang models (Bondi 1961). Averaging any cosmological -- or indeed any physical -- quantities whatsoever with PCP is trivial: after allowing for local fluctuations, the average value is always equal to the currently observed one. This cannot hold in the evolving cosmologies, however, with mere spatial CP. There has been an interesting twist here in the context of the last two decades and the development of inflationary cosmology: generic forms of chaotic inflation tend to produce the inflationary multiverse, which might realize PCP (e.g., Linde, Linde, and Mezhlumian 1994). The measurement problem in this case is also quite complex and not well understood.  
  
Can time be regarded independently of the reference class of objects? For starters, some entities might not be persistent. E.g., galaxies or stars may cease to exist at some point in the future. Physical eschatology suggests that even in formally infinite future of open or flat models, all bound structures, from nucleons to galaxies, including even black holes, have their expiration date (Adams and Laughlin 1997). Galaxies, for instance, will lose all stars through either slingshot evaporation or collapse to the central supermassive black hole at the epoch about $10^{19}$ years after the Big Bang -- a huge timescale for sure, but equally negligible from the point of view of future temporal infinity as the present age of the universe. Therefore, averaging over the entire temporal interval $[0, +\infty)$ might be problematic, since after some critical epoch we lack the entities in question. 

The second problem is that even entirely natural processes might fluctuate on sufficiently large timescales. Consider a well-defined property of the observed universe, for example the baryonic mass fraction of gas in an $L^*$ galaxy like the Milky Way, $f_g$. This goes from 1 in the distant past, before any star formation took place when all baryons were in the gas form, to some value near 0 in the course of future cosmological evolution. Note that it will not formally go to exactly 0 at any given final moment in the future, until other destructive processes, like the proton decay are taken into account on much longer timescales of $10^{34}$ years or more. Suppose that we even restrict the temporal interval to $[0, t_{\rm evap}]$, taking into account the evaporation of stars from galaxies as per results of Adams and Laughlin. What is its mean value, however? We can build models of the rate of gas consumption and its future evolution, but there are multiple unavoidable uncertainties related to such enterprise: of the model, of the underlying theory, of boundary conditions, etc. In particular, it is obvious that gas consumption through star formation will cease at some future epoch (see Section 2 below), but complex feedbacks prevent us from ascertaining when. In addition, new long-term effects which are neglected today will become relevant in the eschatological future -- even after the cessation of conventional star formation, a slow process like accretion of interstellar gas by brown dwarfs will continue and might result in further reduction of $f_g$ (\'Cirkovi\'c 2005). These complications have not been analysed so far in any kind of detail, due to rather poor theoretical understanding of physical eschatology. 

Finally, a potentially huge problem to be encountered in future epochs, in contrast to past epochs, are intentional influences. While living beings have influenced their physical environment in the past as well (e.g., in the \textit{Great Oxygenation Event} about 2.45 Gyr ago; Holland 2006), this pales in comparison to what we could expect in the future, especially in connection with the advanced technological civilizations described by Kardashev's classification (Kardashev 1964, 1985; \'Cirkovi\'c 2015). Even in the particular case of very immature human civilization, introduction of the concept of the \textit{anthropocene} testifies that the intentional influence of technology on our surroundings has become the legitimate topic of discourse in many disciplines (Ruddiman 2013). Extrapolations of these trends onto the timescales of physical eschatology introduces a completely new type of uncertainty with potential to dominate the future values of parameters we wish to average over. 

So, there are multiple problems with trying to average over the cosmological timescale. In spite of all these difficulties, why is the question of temporal typicality interesting and important? There are several interesting arguments advanced in cosmology, future studies, philosophy, and astrobiology which rely on the ``Copernican'' assumption of temporal typicality. The prototype is the ``Doomsday Argument'' in the version of Gott (1993), which follows from the affirmative answer to the question: can we predict the duration of X from X's present age? A particularly interesting example is the recent study of Olson (2017), which advances the argument that a more dangerous early universe increases the number of habitable sites at the universe today. Among the several assumptions used in the argument, the crucial one is of temporal Copernicanism, i.e., \textit{that our particular cosmic time is close to typical}. (Note that temporal Copernicanism is diametrically opposed to {\em chronocentrism}, as formulated e.g., by Fowles 1974 as the belief ``that one's own times are paramount''. The undermining of temporal Copernicanism does not mean any endorsement of chronocentrism, especially not as this is misused in social and political context; we shall return to this point in the concluding section.) If the temporal Copernicanism is undermined, so are arguments such as Olson's. One can go even further and argue that tacit temporal Copernicanism is often responsible for misunderstanding of much of the physical eschatology: we look in hindsight at the evolutionary processes leading to the present moment and privilege them in the entire set of all evolutionary processes occurring at all timescales. While a sceptic can argue that this form of bias is impossible to avoid for temporal observers such as ourselves, this sceptical reasoning does not necessarily end the discussion; we can establish fairly precise theoretical predictions for processes which have not and indeed could not be empirically established so far, such as the Hawking evaporation of black holes, natural formation of positronium, etc. In itself, this temporal bias does not {\em force} us to accept temporal Copernicanism, just as the empirical fact that we have evolved in a galaxy does not force us to assume anything particular about our spatial location within it. Quite to the contrary, our astrobiological research have led us to the concept of the Galactic Habitable Zone, thus refuting the naive Copernican view that our position in the Milky Way is random, unconstrained, or indeed typical (e.g., Gonzalez 2005). 

Therefore, in the rest of this paper, we wish to defend a deflationary thesis that we cannot entertain typicality in time and use it in anthropic (or other) arguments. This, obviously, does not preclude using other kinds of anthropic arguments; if anything, it supports anthropic reasoning by delineating its limits and helping focus on important open questions. The present considerations apply only to the {\em classical cosmological discourse}; averaging in quantum cosmology is an entirely different and complex issue. In Section 2, we discuss technical aspects of temporal averaging and difficulties one encounters in trying to establish temporal typicality. In Section 3, the deflationary view of temporal Copernicanism is elaborated, before recapitulation and discussing some prospects for further work in the concluding Section.

\section{Temporal typicality?}
What does it mean for a quantity, say $A$, which we measure or theoretically consider here, on Earth, to be typical {\em in space}? Clearly, one natural way is to claim that its local value $A(x_0)$ is not very different from the spatial average (mean value):
\begin{equation}
\vert A(\vec{x_0})-\bar{A}\vert < \epsilon,
\end{equation}
where  $\vec{x_0}$ is our spatial location in a well-defined coordinate system and $\epsilon$ is a small positive real number. The mean value is obtained by averaging over the relevant volume:
\begin{equation}
\bar{A}=\lim_{V\rightarrow V_{\rm max}}\frac{1}{V}\displaystyle\int\limits_V A(\vec{x})d^3\vec{x},
\end{equation}
where $V$ is the relevant volume limited by:
\begin{equation}
V_{\rm max}=\left\{ \begin{array}{cr} \infty, & \textrm{open and flat Friedmann models} \\ V_{\rm tot}, & \textrm{closed Friedmann models}\end{array}\right.
\end{equation}
which {\em subsumes} CP in the very definition of the Friedmann models. The limiting process in (2) is uncontroversial as long as $A(x)$ is well-behaved, which is the case for all astrophysical quantities of interest. Alternatively, one could use the median of the distribution of $A$. For continuous data, the median value $A(m)$ is the value of the function for argument m such that the area (or integral) of the data to one side of the point is equal to the area on the other side: 
\begin{equation}
\frac{\displaystyle\int_m^{\lambda_{\rm max}} A(\lambda)d\lambda}{\displaystyle\int_{\lambda_{\rm min}}^{\lambda_{\rm max}}A(\lambda)d\lambda}=\frac{1}{2}.
\end{equation}
Astrophysical quantities dependent on cosmic time, like the CMB temperature or the star-formation density in a particular galaxy or a type of galaxies, which can adequately be represented by continuous real functions, can play the role of A here. In each case, however, there is a host of requirements which need to be satisfied in order for the mathematical machinery to work. Even the more general Lebesgue-Stieltjes integration is not necessarily well-defined in the integrals here (e.g., Shilov and Gurevich 1978). The reason for this is, obviously, the fact that CP is valid only on large scales, strictly in the lengthscale $\lambda\rightarrow \infty$ limit\footnote{Strictly speaking, the validity of the CP is only testable within the causal horizon, and there may very well be large deviations from the FLRW metric beyond the Hubble volume. This, however, has no practical consequence for this argument.}, and in the real universe we have all sorts of structures to deal with. Local structures cause deviations from the FLRW metric which often {\em cannot be treated as small perturbations} to the ``regular'' FLRW background. 

Therefore, it has been acknowledged recently that even the spatial averaging is far from being trivial. Many new studies testify on the increased interest in this issue (e.g., Coley 2010; Ellis 2011; Clarkson et al. 2011; Ka\v{s}par and Sv\'itek 2014). In a prestigious review (Clarkson et al. 2011, p. 2):

\begin{quotation}
\textit{The Universe may well be statistically homogeneous and isotropic above a certain scale, but on smaller scales it is highly inhomogeneous, quite unlike an FLRW Universe. General relativity is a theory in which spacetime itself is the dynamical field with no external reference space. Yet it is ubiquitous in cosmology to talk of a `background' which is exactly homogeneous and isotropic, on which galaxies and structure exist as perturbations. Is this the same as starting with a more detailed truly inhomogeneous metric of spacetime, and progressively smoothing it -- probably by a non-covariant process -- until we get to this background?}
\end{quotation}

This problem gave rise, in the last decade, to a debate around the importance of back-reactions, i.e. the effect of inhomogeneities on the average evolution of the universe. In essence, doubts have been raised with regard to the standard calculations of the non linear growth of density perturbations, which is usually treated with a Newtonian approximation decoupled from the average background expansion of a uniform universe. If true, this objection would put into question the interpretation of several cosmological observations, including the accelerated expansion and its dominant interpretation in terms of a dark energy component. As of now, the debate is not settled (see, e.g. Green and Wald 2014 and Bucher et al. 2015 for two opposing takes on the subject). We note that the problem of spatial averages cannot be entirely disentangled from the question of time in cosmology, since there have been proposals that the operation of synchronizing clocks and defining cosmic time is ill-defined in an inhomogeneous universe (Wiltshire 2009).

Also (Clarkson et al. 2011, p. 5):
\begin{quotation}
\textit{Averaging is in some respects a fitting process, but does not necessarily correspond to any actual observational procedure. Can one propose an average model of the Universe based on the past-null cone? [...] it does not make sense to simply average today's state of the Universe with what it was like in the past. One would expect any averaging operation to leave the background invariant, and it is not obvious that this can happen for FLRW somehow averaged on its light cones. So averaging based on observations would need to involve comparing the Universe today with earlier times by the use of dynamical equations relating variables at these different times: a very model dependent procedure, and not `averaging' in a normal sense.}
\end{quotation}

The last conclusion is very important from the point of view of the deflationary account of the typicality in time, which will become clearer in the next section. What would be the ``background invariant'' for averaging over time? It is doubtful that such concept is intelligible at all in the temporal case. 

In other words, consider the standard expression for spatial average (e.g., Weinberg 2008): 
\begin{equation}
\bar{A}(\vec{x},t)\vert_{\mathfrak{I}}=\frac{1}{V_\mathfrak{I}}\displaystyle\int\limits_\mathfrak{I}A(\vec{x},t)\sqrt{\det h}\; d^3\vec{x},
\end{equation}
where $\vert_{\mathfrak{I}}$ denotes averaging over a region $V_\mathfrak{I}$  of a spatial hypersurface $\mathfrak{I}$.
Suppose, for instance, that we try to define temporal average by analogy with (2) as:
\begin{equation}
\langle A\rangle=\lim_{t\rightarrow t_{\rm max}}\frac{1}{t}\displaystyle\int_0^t A(x)dx,
\end{equation}
\begin{equation}
t_{\rm max}=\left\{ \begin{array}{cr} \infty, & \textrm{open and flat Friedmann models} \\ t_{\rm tot}, & \textrm{closed Friedmann models}\end{array}\right.
\end{equation}
The obvious problem is that the limiting process, $t\rightarrow t_{\rm max}$ is not necessarily legitimate in either mathematical or physical sense. The functions under consideration need not be regular, not to mention continuous in the temporal realm, even if they have intelligible and well-defined meaning at all epochs. Even those which are expected to behave reasonably well might be either sensitive on boundary conditions in an unpredictable way, or simply go quickly to zero, driving all averages to vanish asymptotically (in the case of open-future cosmologies to which the post-1998 ``new standard'' cosmology seemingly belongs).

Let us consider the following example of actual interest for astrophysics and astrobiology which can serve as a toy model to highlight the problem. Consider the question: how long will the current epoch of active star-formation in the universe (dubbed the \textit{stelliferous} era by Adams and Laughlin 1997) last? Clearly, the question is of significant import for understanding of the evolution of stellar populations and galaxies themselves. It is also of paramount importance for physical eschatology as the ``cosmology of the future'' -- but also for astrobiology as long as we accept that physical conditions such as those encountered on Earth are necessary for the emergence of life and intelligence. 

Star formation histories of spiral galaxies are determined by the interplay between incorporation of baryons into collapsed objects (stars, stellar remnants and smaller objects, like planets, comets or dust grains) and return of baryons into diffuse state (gaseous clouds and the intercloud medium). The latter process can be two-fold: (i) mass return from stars to the interstellar medium (henceforth ISM) through stellar winds, planetary nebulae, novae and supernovae, which happens at the local level; and (ii) net global infall of baryons from outside of the disk (if any). 

In a seminal study, Larson, Tinsley, and Caldwell (1980) considered the timescale for gas exhaustion from spiral disks due to star formation (following the program outlined in the early study of Roberts 1963), in an attempt to justify the bold hypothesis that S0 galaxies may be disk galaxies that lost their gas-rich envelopes at an early stage and consumed their remaining gas by quick star formation. They found that the appropriate Roberts' timescales for removal of ISM in spiral disks are in most cases rather short in comparison to the Hubble time -- a conclusion which in itself suggests a conflict with temporal Copernicanism. In the roughest approximation, we can define the timescale of gas consumption:
\begin{equation}
\tau_{\rm gas}=\left\vert \frac{\Sigma_{\rm gas}}{\Sigma_{\rm sfr}}\right\vert,
\end{equation}
where $\Sigma_{\rm gas}$ is the gas surface density in units [$\Sigma_{\rm gas}$]$=M_\odot$ pc$^{-2}$  and $\Sigma_{\rm sfr}$ the star formation rate, $\Sigma_{\rm sfr}=d\Sigma_{\rm gas} /dt=\Sigma_{\rm gas}$  in units [$\Sigma_{\rm sfr}$]$=M_\odot$ Gyr$^{-1}$pc$^{-2}$  The empirical data in a sample of 61 nearby spirals of Kennicutt (1998) are shown in Figure 1, clearly showing preference for short ($< 3$ Gyr) timescales and the conclusion that we are living in a special cosmological epoch, contrary to temporal Copernicanism. 

\begin{figure}[th!]
\center{\includegraphics{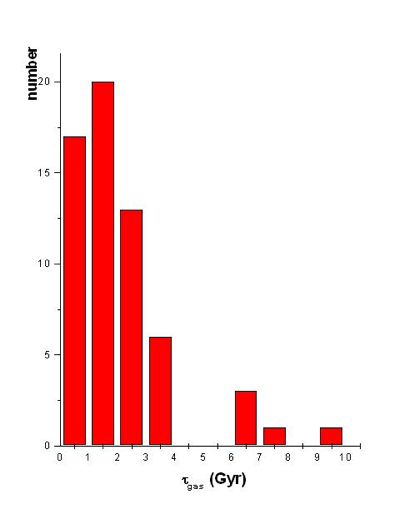}}
\caption{The distribution of baseline gas consumption times based on the current rate of star formation in the sample of Kennicutt (1998).}
\end{figure}

The major development in the field of the global star formation was the realization that there exists a star formation {\em threshold} at finite gas density or disk surface density (Martin and Kennicutt 2001, and references therein). Empirical threshold gas surface density seems about $6 M_\odot$pc$^{-2}$. Now, it obviously cannot be the whole story, since we are aware of additional processes, notably recycling of gas and nonlinear dependence of the star formation rate on gas density (Schmidt's Law), not to mention possible massive infall at late epochs. In general, one needs to integrate the equation
\begin{equation}
\frac{d\Sigma_{\rm gas}}{dt}=-\left[1-r(t)\right]\Sigma_{\rm sfr}+I(t),
\end{equation}
where $r(t)$ is the gas return fraction from stars to ISM, integrated over the entire population of stars. For the classical Miller-Scalo initial mass function, this value today is $r(t_0)=0.42$. $I(t)$ is the gaseous infall rate; for the sake of simplicity, we can take the Gaussian form (e.g., Prantzos and Silk 1998):
\begin{equation}
I(t)=\frac{\mu}{\sqrt{2\pi}\sigma}\exp{\left[-\frac{(t-\tau_{\rm inf})^2}{2\sigma^2}\right]}.
\end{equation}
Here $\mu$ is the normalizing mass scale for the infall, and $\sigma$  and $\tau_{\rm inf}$ are the infall temporal width and the characteristic epoch both with dimensions of time. Prantzos and Silk use fiducial values for these parameters as $\sigma=\tau_{\rm inf}=5$ Gyr. These are constrained by the present-day infall $I_0\equiv I(t_0)$  as:
\begin{equation}
\mu=I_0\sqrt{2\pi}\sigma \exp{\left[\frac{(t_0-\tau_{\rm inf})^2}{2\sigma^2}\right]}.
\end{equation}
For Schmidt's Law we may use the usual {\em ansatz}  $\Sigma_{\rm sfr}(t) = A [\Sigma_{\rm gas}(t)]^n$ where $A$ is the conversion function independent of time (although it may vary with the galactocentric radius, as indicated in several theoretical studies). For the index of Schmidt's Law, we use the value of $n=1.3\pm 0.2$, which agrees with the Kennicutt (1998) study. 
All in all, the impact of these complications is shown in Figure 2. 

\begin{figure}[th!]
\center{\includegraphics{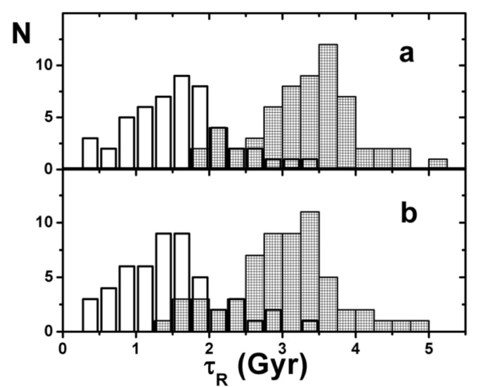}}
\caption{Predictions for the gas exhaustion timescales (Roberts' times) in a model with Schmidt's Law, Gaussian infall, and thresholds taken into account. In panel a recycling is taken into account as well, while it is neglected in the panel b (the difference is almost negligible). Hollow rectangles correspond to the fixed gas surface density threshold of $6 M_\odot$pc$^{-2}$, and the shaded ones to the half that value.}
\end{figure}

This is obviously a toy model, but it highlights some of the important features of limits on the stelliferous era. The distribution of durations changes in shape with taking additional physics into account, but does not show a substantial increase in the median timescale. If anything, the median slightly {\em decreases}, since the star-formation thresholds are very efficient in arresting active star formation, as is seen in well-studied individual galaxies; in one of prototypical normal SA(s)c spirals, NGC 4254, it is easy to calculate that {\em even in the absence of thresholds} gas density will fall below 1 Solar mass per pc$^2$ in about 10 Gyr. Of course, one could try to play with different parameters like the infall mass scale or the ``true'' index of Schmidt's Law in order to select parts of the parameter space corresponding to an {\em increase} in the duration of future star formation. This, however, should not occlude the two main points of relevance for the present study: (i) that part of the parameter space is {\em small} and requires fine tuning, thus returning us to the very same difficulties temporal Copernicanism purports to resolve; and (ii) the average star formation rate taken in the sense of Eq. (6) is {\em zero}. Therefore, our observation that we are living in the epoch of active -- though declining -- star formation in ISM of the Milky Way and other normal galaxies is, strictly speaking, in conflict with the temporal Copernicanism. 

One could go even further and note that there are stars -- like M-dwarfs -- which live orders of magnitude longer than the Hubble time staying in an essentially unchanging state on the Main Sequence for up to $10^{13}$ years (e.g., Laughlin, Bodenheimer, and Adams 1997). Contemporary astrobiology reveals that they have planets and at least potentially habitable Earth-like planets (Anglada-Escud\'e et al. 2016; Astudillo-Defru et al. 2017). Searches for biosignatures on planets around those extremely long-lived M-dwarfs are currently under way. Note how much longer their lifetimes are than even the most optimistic estimates of the future duration of the stelliferous era. This means that there can be no ``saving clause'' for temporal Copernicans which would necessitate observers to be located in the stelliferous era (and hence in an atypical cosmic time). Even in the absence of large-scale astroengineering feats of advanced technological civilizations (which are perfectly in accordance with the laws of physics) which could extend the lifetimes of observers and their communities still some orders of magnitude more into the post-stelliferous eras of the universe, one expects observers to arise and perhaps go extinct long after conventional star formation ceases. 

\section{Deflationary view and Copernicanism}
In a more relaxed and general sense, the problem with averaging over large time spans in nonlinear dynamical systems is not limited to cosmology. Similar situation might occur in evolutionary biology, with respect to the contentious issue of the size of the genome space sampled by evolution on Earth so far (Dryden, Thomson, and White 2008; McLeish 2015). How to assess whether the current state of the terrestrial biosphere is indeed typical with respect to the genomic diversity? Usually it is just {\em assumed}, but the assumption is impossible to falsify, since it is unclear whether the question is actually well-posed: how about future genomes created artificially by humans (or posthumans) which could be designed, at least in principle, to be arbitrarily distant from the parent population in the genome space? These will also be part of the terrestrial biosphere on the most reasonable construals: however, accounting for their diversity is not only impossible at present, but also will immensely bias the quantitative outcome. The genome space is huge -- resulting from combinatorics, essentially -- and we can empirically sample only a small part of that already minuscule part which was sampled in the course of the 4.1 Gyr of the evolution of terrestrial life. This would correspond to the standard view (or ``received wisdom'') on this matter. Actually, whether evolution on Earth has sampled a minuscule or a significant part of the viable genome space has been the subject of some contention recently (e.g., Dryden, Thomson, and White 2008; McLeish 2015; Powell and Mariscal 2015).  And, just as in the cosmological examples, the absence of meaningful averages makes any conclusion about typicality of the present-day biospheric genome space untestable and hand-waving. It is entirely plausible that there are further such examples. 

Confronted with all these problems, it is only intellectually honest and rational to admit the difficulties and adopt a more modest, deflationary account of typicality in time: {\em we cannot really say whether we are living in a typical cosmological epoch and we should therefore refrain from assuming our typicality in this respect and deriving any conclusion from such an assumption}. 

In brief, at our present level of knowledge there is no way to tell whether evolution of sufficiently complex and sufficiently nonlinear world is sufficiently ergodic for whatever mean values are required to be well-defined at all. And even if they were well-defined, the practical task of computing them is far beyond our present capacities, requiring predictive powers unlike anything we are dealing with in science. Therefore, in contrast to some of the other ideas about typicality, the typicality of present epoch is extremely difficult and possibly impossible to achieve even in principle, making the relevant Copernican assumption more metaphysical and less scientific than any other typicality assumption. Again, the assumption that the Sun is a typical star is rather easy to theoretically specify and quantify, as well as empirically check now by observing a large sample of stars, and it was in principle verifiable even in the time of Copernicus and Galileo. In sharp contrast, the assumption that we are living in a typical cosmological epoch is {\em both theoretically confusing and empirically untestable}; we are not sure how to proceed, even in theory. 

In fact, at a very basic level, there is a fundamental limitation in our ability to predict the future evolution of the universe: if the cosmological constant (or a more generic dark energy component) is non-zero (as in the currently favoured cosmological model) than it is straightforward to show that {\em no set of cosmological observations can unambiguously determine the ultimate destiny of the universe} (Krauss \& Turner 1999). This is often misinterpreted as a sort of technical difficulty; in fact, it is a limitation so fundamental that it makes any kind of long-term integration such as in Eq. (6) above meaningless. In such a position, the correct course is to accept that -- until further insights come in -- we cannot truly argue that our temporal location is typical. This would reject arguments such as Gott's (who in fact tacitly admits it, strangely enough, by failing to acknowledge the failure of PCP among listing the successes of Copernicanism in the introductory part of his 1993 article) or Olsen's. 

As further example, consider the temporal distribution of habitable planets in the Milky Way, as calculated by Lineweaver (2001) in the pioneering study of the topic, of enormous importance for astrobiology and SETI studies (for subsequent more precise elaborations, see Lineweaver, Fenner, and Gibson 2004; Behroozi and Peeples 2015; Zackrisson et al. 2016).  Lineweaver has established that the formation of Earth-like planets started somewhat more than 9 Gyr ago and their median age is $6.4\pm 0.9$ Gyr. So, we have obtained a temporal distribution of ages of a well-defined class of objects with a definite fixed start, reaching maximum, then declining to the present-day value and extending for an indefinite amount of time into the future. The study itself does not say anything about the future distribution (since it deals with other issues of relevance for astrobiology), but it is clear that the future extension does exist. More generally, recent studies have started addressing the problem of the overall habitability of the universe in time (e.g. Dayal et al. 2016, Loeb et al. 2016). This kind of investigation is still in its infancy and its conclusions are very uncertain. However, there are no compelling reasons to expect that the probability for the appearance of life in the universe is uniform in time, and in fact there are quite more arguments to the contrary. This very fact clashes with the presumption that we are observing a typical epoch of cosmic history.

The unwarranted assumption of uniformity for the temporal distribution of intelligent observers in the universe has also strong consequences when assessing the chances of success of SETI (or, more broadly, the search for signatures of technological species, or ``technosignatures''). In fact, most past studies in the field adopted, explicitly or not, an underlying presumption of stationarity for the appearance of life over cosmic history: one notable example of this approach is the use of the Drake equation (Drake 1965) to estimate the number of communicating species present in the universe, in which any time dependence of the relevant astrophysical or biological factors is neglected. When evolutionary effects are properly taken into account, however, one can derive very different estimates of the number of detectable technosignatures in the universe (\'Cirkovi\'c 2004, Balbi 2018). Similarly, relaxing the assumption of our temporal typicality can render moot many discussions on the so-called Fermi paradox, i.e. the apparently puzzling fact that we have not encountered any evidence of advanced civilizations in the universe. 

\section{Discussion}
The problem of temporal (un)typicality presents an excellent example of a problem in astrophysics/physical cosmology with an important philosophical component. One of the major methods of modern analytic philosophy, semantic analysis of terms such as ``average'' or ``universal'', obviously plays an important role in resolving -- or even getting better insight into -- the issue. While spatial averaging brings no new unknowable elements, and the uncertainty hinges on rather fine details like the back-reaction of structure formation which is at least potentially observable, temporal averaging brings radically new and at least in part unknowable elements, since the future is not similar to the past.

An uncharitable mischaracterization of the present argument suggests that since the future is uncertain, we could never hope to establish whether creatures like us could exist in other epochs. Clearly, the deflationary view of typicality suggests much more -- even without the inherent ambiguity present in the word ``uncertain''. As per President Bush's funny quote above, we can agree that the past is over -- but we lack even vague and provisional consensus about the future in most cases, so that the conclusion about present being typical is as irrational as if somebody stopped watching a football match at half-time on the pretext that the current score is typical enough. (And in the cosmological case, it could be not 50\%, but up to $10^{-80}$ of the total match!) In some particular cosmological contexts, we could go much further along the deflationary path: Hartle and Srednicki (2007) present a convincing case that one cannot falsify a theory on account of it predicting that we are not typical. Together with what has been said above, we may even argue that typicality, in general, provides very little inferential value for a model -- it may be a preferred prior, not a model selection tool. Even as a prior, it is hardly a very robust one; its fragility depends on the specific problem situation, but it needs to be taken into account in any specific discussion and not swept under the rug. 

As a further example of how a typicality assumption can lead to conundrums and to dubious model selection criteria, we can mention the flurry of recent cosmological discussions about the so-called ``Boltzmann brains'' problem (Dyson et al. 2002): in short, the argument is that if our universe lasts forever and reaches an asymptotical De Sitter phase (as in the currently favoured cosmological constant scenario), one might expect random fluctuations away from thermodynamical equilibrium (i.e. from higher to lower entropy) to give rise, in the future, to conscious observers that would vastly outnumber those arising as a result of ``ordinary'' thermodynamic processes (i.e., from lower to higher entropy). The argument goes on by using our own untypicality (as long as we can be sure that we are not a Boltzmann brain) as a ``proof'' that there is something wrong with the current cosmological model (or with extensions of it, such as the eternal inflation scenario). Our untypicality with respect to hypothetic Boltzmann brains has also been used to argue that the universe is likely to end in a finite time (Page, 2008). Note that, although the Boltzmann brain argument is not usually phrased as stemming explicitly from a temporal typicality assumption, in fact it is strictly linked to that, since it takes as puzzling the fact that we are observing the universe now, given that most observers in the universe should be temporally located in the extremely far future of a very long-lived universe. If one gives up a naïve assumption of our temporal typicality (or of typicality {\em in general}), however, the argument loses much of its strength.

The present view is related to the suggestion of Ben\'etreau-Dupin (2015) that problems formulated in terms of Bayesian induction, like Gott's (1993) version of the traditional ``doomsday argument'' or various fine-tuning problems in cosmology, could be resolved by an imprecise, ``blurred out'' approach. The deflationary view, when applied to probabilistic problems, could be reformulated as implication that our ignorance about future evolution implies a multiplicity of credence functions. This degeneracy is a particular instance of the general failure of induction in the domain of the futures studies/physical eschatology: 
\begin{quotation}
\textit{These cosmic puzzles show that, in the absence of an adequate representation of ignorance or indifference, a logic of induction will inevitably yield unwarranted results. Our usual methods of Bayesian induction are ill equipped to allow us to address either puzzle. I have shown that the imprecise credence framework allows us to treat both arguments in a way that avoids their undesirable conclusions. The imprecise model rests on Bayesian methods, but it is expressively richer than the usual Bayesian approach that only deals with single probability distributions.} (Ben\'etreau-Dupin 2015, pp. 889-90.)
\end{quotation}

Note that this does not apply solely to ``cosmological puzzles'' -- even the future of Earth or of any other complex system manifests a similar failure of simplistic assumptions of temporal Copernicanism. (Although, in any restricted context, the degree of approximation in assuming that the system is isolated becomes a source of noise over and above the intrinsic uncertainties in the evolutionary dynamics.)

The deflationary view of the typicality in time suggested here needs not be worrying for committed Copernicans: there are so many other parameters and instances in which the Copernican principle is a trustworthy guide. If anything, anti-Copernican attempts such as the controversial ``rare Earth'' hypothesis of Ward and Brownlee (2000) erroneously predicted that there are very many ``hot Jupiters'' in the Milky Way, and also that elliptical galaxies should not be considered habitable (see Dayal et al. 2015 for a persuasive argument to the contrary). On the other hand, the deflationary view does impact those particular arguments which {\em posit} our temporal location as typical, such as Olson's argument discussed above or the doomsday argument of Gott. 

We note that sensible applications of Copernicanism in time are in principle still possible by focusing on subsets of the history of the universe where we might expect a reasonably uniform behavior of some variable quantity relevant to the problem under investigation. For example, when discussing the chances that life evolved on other planets, one might restrict the time interval only to the 'stelliferous' era, or some chunk of it. But this is precisely the main caveat of our work: there is no straightforward way in which our typicality in time can be {\em assumed}, without accompanying the assumption with some knowledge of the temporal evolution of key factors involved in the habitability of the universe and in the appearance of life. So, at the very least, our analysis highlights the necessity of further work to understand the complicated variation in time of the conditions that make the presence of observers possible. 

There is another important difference between the legitimate applications of Copernicanism, for instance, in spatial domain, and the illegitimate temporal Copernicanism; this difference belongs to the domain of epistemology. We are in principle free to travel through space, and spatial boundary conditions for our theoretical models are subject to empirical verification (again, at least in principle and subject to relativistic constrains). In the temporal domain we are blocked -- barring closed timelike curves or other forms of two-way time travel -- from direct inspection of boundary conditions. In the course of temporal evolution, coarse-graining erases accessible information (not information {\em sub specie aeternitatis}, as even Stephen Hawking admitted in 2004!). Obviously, this is the reason why some manifestly scientific questions cannot be answered in practice and are often confused with metaphysical issues (e.g., why the apparent sizes of the Sun and the Moon are so similar? why Venus has no large moons? why life on Earth manifests homochirality? etc.). There is a chance, smaller than usually assumed, that further work on rare and tenuous information reaching us from the distant past will one day enable answering these such questions; this applies to the future as well, however, since there is a small chance that one day we will reach detailed and precise models of future evolution of various complex systems. In contrast to implicit claims of temporal Copernicans, the macroscopic symmetry is restored to epistemology -- future temporal boundary conditions are at least as poorly understood as the distant past temporal boundary conditions (see also Price 1996). 

What can we do to improve our understanding of our temporal position? Obviously, lacking empirical transtemporal perspective, what is required is to have better theoretical insight into the long-term evolution of various physical systems. In particular, the nascent discipline of physical eschatology is of paramount importance for any averaging over cosmological time. There is a wealth of interesting results to be found in studying the future of the universe, especially when massive numerical simulations are applied to it, what has not been the case so far. To these studies, one should add at least cursory examination of the impact of advanced technological civilizations on their physical surrounding, which gives rise to effects which are certainly non-negligible in the cosmological future. Research programmes in this area offer promise of the next significant multidisciplinary synthesis: the one of cosmology with astrobiology and SETI studies. Like the previous grand synthesis of this kind -- the one of cosmology and particle physics which took place mainly in 1980s -- this one is likely to bring fruit far in excess of anything envisioned at the time of its conception. 

\subsection*{Acknowledgement}{The authors acknowledge Mark Walker and Marko Stalevski for useful comments.}

\end{document}